\documentclass[10pt, a4paper]{article}

\usepackage[]{lrec-coling2024} % this is the new style
\usepackage{microtype}
\usepackage{array,multirow}
\usepackage{xcolor}
\usepackage{booktabs}
\usepackage{epigraph}
\usepackage{enumitem}
\usepackage{etoolbox}
\usepackage{subfig}
\usepackage{hyperref}
\usepackage{balance}

\definecolor{green}{rgb}{0.0, 0.5, 0.0}
\definecolor{gray}{rgb}{0.33, 0.41, 0.47}

\usepackage{orcidlink}
\newcommand{\CTorcid}{\orcidlink{0000-0001-6976-3258}}
\newcommand{\FSorcid}{\orcidlink{0000-0001-9104-2860}}
\newcommand{\OGorcid}{\orcidlink{0000-0002-9420-9854}}
\newcommand{\FCorcid}{\orcidlink{0000-0003-0016-4278}}

\usepackage{scalerel,graphicx,xparse}
\NewDocumentCommand\emojiA{}{
    \includegraphics[scale=0.06]{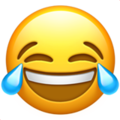}
}
\NewDocumentCommand\emojiB{}{
    \includegraphics[scale=0.06]{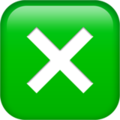}
}
\NewDocumentCommand\emojiC{}{
    \includegraphics[scale=0.06]{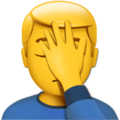}
}
\NewDocumentCommand\emojiD{}{
    \includegraphics[scale=0.06]{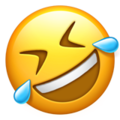}
}
\NewDocumentCommand\emojiE{}{
    \includegraphics[scale=0.06]{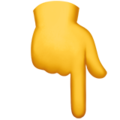}
}
\NewDocumentCommand\emojiF{}{
    \includegraphics[scale=0.06]{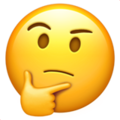}
}
\NewDocumentCommand\emojiG{}{
    \includegraphics[scale=0.06]{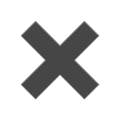}
}
\NewDocumentCommand\emojiH{}{
    \includegraphics[scale=0.06]{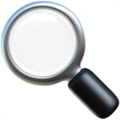}
}
\NewDocumentCommand\emojiJ{}{
    \includegraphics[scale=0.06]{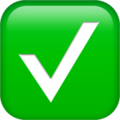}
}
\NewDocumentCommand\emojiK{}{
    \includegraphics[scale=0.06]{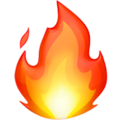}
}
\NewDocumentCommand\emojiL{}{
    \includegraphics[scale=0.06]{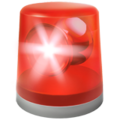}
}
\NewDocumentCommand\emojiM{}{
    \includegraphics[scale=0.06]{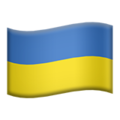}
}
\NewDocumentCommand\emojiN{}{
    \includegraphics[scale=0.06]{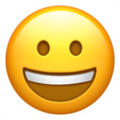}
}
\NewDocumentCommand\emojiO{}{
    \includegraphics[scale=0.06]{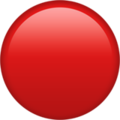}
}

\title{MiDe22: An Annotated Multi-Event Tweet Dataset for Misinformation Detection}

\name{Cagri Toraman$^1$*\CTorcid, Oguzhan Ozcelik$^{2,3}$\OGorcid, Furkan \c{S}ahinu\c{c}$^4$*\thanks{*Work partially done in Aselsan, Ankara, Turkey.}\FSorcid, Fazli Can$^2$\FCorcid}

\address{$^1$Department of Computer Engineering, Middle East Technical University, Ankara, Turkey \\
$^2$Department of Computer Engineering, Bilkent University, Ankara, Turkey $^3$Aselsan, Ankara, Turkey\\
$^4$Ubiquitous Knowledge Processing Lab (UKP Lab), Technical University of Darmstadt \\
    \texttt{ctoraman@ceng.metu.edu.tr} \\
    \texttt{oguzhan.ozcelik@bilkent.edu.tr} \\
    \texttt{furkan.sahinuc@tu-darmstadt.de} \\
    \texttt{canf@cs.bilkent.edu.tr} \\
}

\abstract{
The rapid dissemination of misinformation through online social networks poses a pressing issue with harmful consequences jeopardizing human health, public safety, democracy, and the economy; therefore, urgent action is required to address this problem. In this study, we construct a new human-annotated dataset, called \texttt{MiDe22}, having 5,284 English and 5,064 Turkish tweets with their misinformation labels for several recent events between 2020 and 2022, including the Russia-Ukraine war, COVID-19 pandemic, and Refugees. The dataset includes user engagements with the tweets in terms of likes, replies, retweets, and quotes. We also provide a detailed data analysis with descriptive statistics and the experimental results of a benchmark evaluation for misinformation detection.
 \\ \newline \Keywords{Human-annotation, Misinformation detection, Multi-event dataset, Tweet} }

\begin{document}

\maketitleabstract

\section{Introduction}

With the growth of online social networks, people develop new behaviors and trends. An example is the amount of news consumed in these networks, and eventually the phrase ``social media'' is coined. However, considering their popularity and easy accessibility, it is inevitable to observe different kinds of content in social media platforms; e.g information manipulations, fake news, and misinformation/disinformation spread\footnote{We use misinformation as an umbrella term that refers to all instances where information have falsehoods.}. Twitter (rebranding to X since July 2023) is one of the platforms where misinformation can be widely spread as observed in the U.S. Elections \cite{Grinberg:2019}, so that ``fake news'' became the Word of the Year in 2017 \cite{Collins:2017}. 

Misinformation is spread in many domains including but not limited to health, politics, and disasters. Once misinformation is spread, the consequences can be devastating \cite{Islam:2020,Reuters:2022}. For instance, many people died because of false rumors that claim that the cure for COVID-19 is drinking methanol \cite{Islam:2020}. Another example is that Ukraine sought an emergency order from the International Court of Justice due to the false claims of genocide against Russian speakers in Ukraine \cite{Reuters:2022}. Considering the importance of misinformation spread in society and the ugly truth of unavoidable diffusion and beliefs, misinformation detection becomes a critical task that requires advanced methods and datasets. 

\begin{figure*}[t]
    \centering
    \includegraphics[scale=0.50]{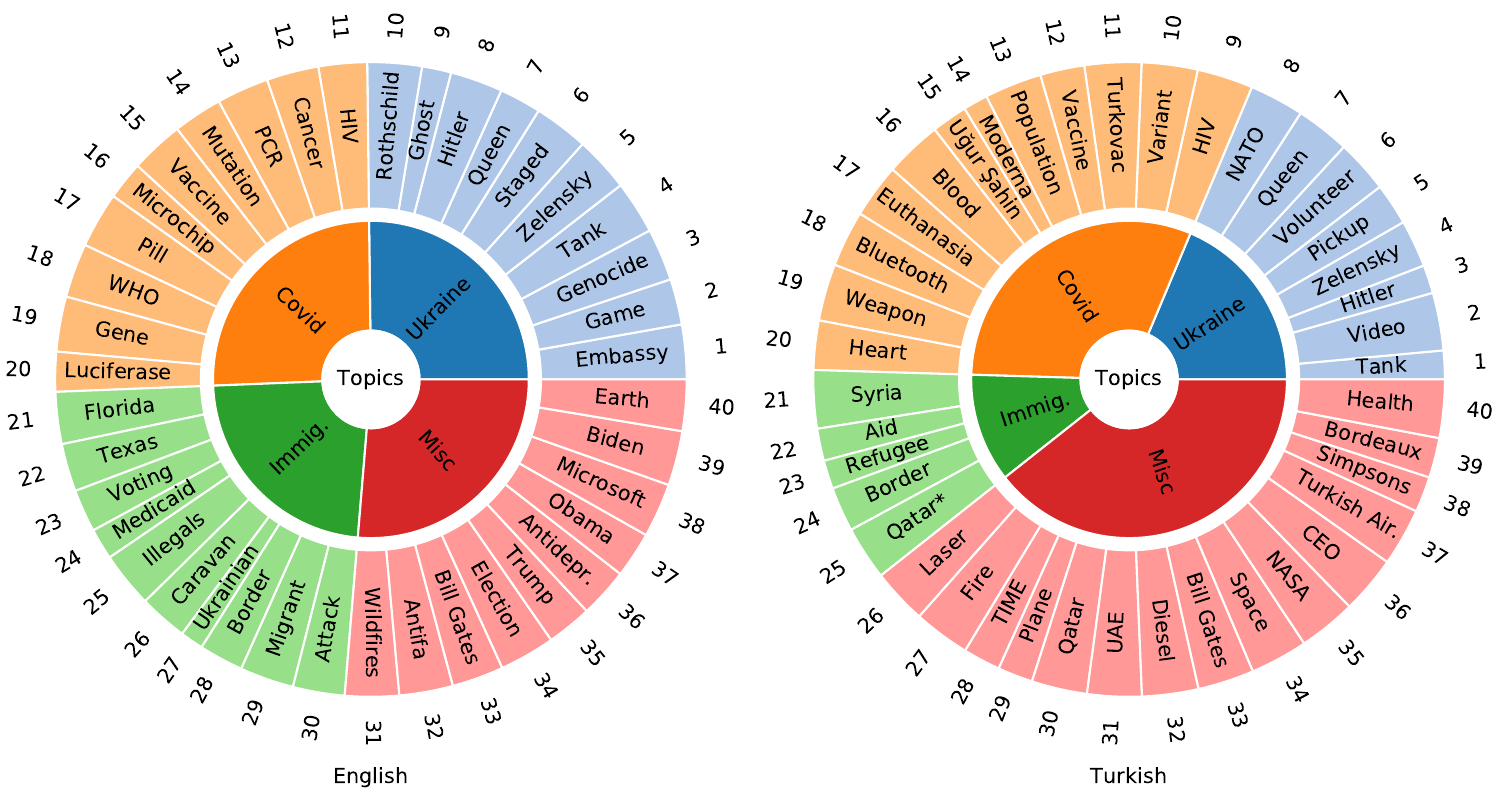}
    \caption{The topics (inner circle) and events (outer circle) in \texttt{MiDe22} for English (left) and Turkish (right). The areas are proportional to the number of tweets they have.}
    \label{fig:composition}
\end{figure*}

A straightforward solution for misinformation is to avoid the spread in advance. However, people can be biased to change their beliefs even if corrections exist, and the attempts to correct falsehoods may not avoid its spread and even sometimes help its diffusion \cite{Nyhan:2010}. Moreover, targeted advertising to increase user engagement can help misinformation spread, which may be a source of revenue for social media platforms \cite{Neumann:2022}.

We have four main observations on existing social media collections for misinformation detection. Although they mostly cover a limited number of topics \cite{Ma:2017}, these topics remain too high-level to provide an opportunity to systematically examine which type of incidents trigger the misinformation spread. The availability of fine-grained event-specific information can play a significant role in capturing different user behaviors for detecting and preventing misinformation. Furthermore, the existing datasets focus on widely used languages such as English \cite{DUlizia:2021}, while they are very limited for low-resource languages. Lastly, user engagements (like, reply, retweet, and quote) and media elements (image and video) in false tweets can be useful to analyze different types of information diffusion and detection methods (e.g. multimodal), but not all types are always included in the datasets. 

In order to bridge these gaps, we present an annotated multi-event tweet dataset for \textbf{Mi}sinformation \textbf{De}tection under several recent events from 2020 to 20\textbf{22}, called \texttt{MiDe22}, including English and Turkish tweets with four types of user engagements and they are likes, replies, retweets, and quotes.

\subsection{Dataset Contents}
The \texttt{MiDe22} dataset\footnote{\label{note1}The dataset and all other related documents can be accessed at \href{https://github.com/metunlp/MiDe22}{https://github.com/metunlp/MiDe22}} consists of three parts: (i) Topics and Events, (ii) Tweets, and (iii) Engagements. Each part exists for both English (\texttt{MiDe22-EN}) and Turkish (\texttt{MiDe22-TR}).

\noindent \textbf{Topics and Events.} We consider the issues occupying the world's agenda in recent years as the topics of our dataset. Then, we extract the significant events with the highest spread of misinformation. Figure \ref{fig:composition} presents an overview of the structure of our dataset. The inner circle indicates the COVID-19 pandemic, the 2022 War between Russia and Ukraine, Refugees (Immigration), and Miscellaneous events that are not categorized under the previous topics. Overall, these topics contain 40 newsworthy events in the outer circles of the figure. We also provide the event titles along with their topics online\footref{note1}. 

Note that we prefer well-known recent events for both languages. The reason is that some misinformation events can be global and observed in several countries, such as ``COVID-19 vaccines contain Human Immunodeficiency Virus (HIV)''. These common events can provide an opportunity to inspect how misinformation is spread in different languages. On the other hand, there are local events that have influence in specific regions. The details on the events are given in Section \ref{sec:data_crawling}.

\noindent \textbf{Tweets.} The dataset has tweets related to the events. The crawling process is explained in Section \ref{sec:data_crawling}. Each tweet is labeled according to three classes: False information, True information, and Other. The Other class includes tweets that cannot be categorized under false and true information. The annotation process is explained in Section \ref{sec:data_annotation}.

\noindent \textbf{User Engagements and Media.} We provide the user engagements with all tweets. Separate engagement splits are provided in the types of like, reply, retweet and quote. We also provide media elements in our dataset, i.e. image and video if they exist in the tweets. 

\subsection{Contributions}
Our contribution involves the development of a novel tweet dataset for misinformation detection in two languages with various topics and user engagements. The languages are a widely used language: English, and a low-resource language: Turkish. The topics of the dataset cover several recent events, such as the 2022 Russia-Ukraine War and the COVID-19 pandemic. The dataset includes the user engagements with all tweets in terms of likes, replies, retweets, and quotes. It can be used in many studies such as misinformation, event, and topic detection. Additionally, we conduct experiments to provide initial baseline scores from different model families, e.g., bag-of-words, neural, and transformer-based models. Apart from demonstrating the quality and utility of our dataset, these baselines also provide a benchmark for researchers to compare against and further enhance their developments. The variety of baseline models is rich enough to perform statistical tests and interpret the results properly.

\begin{table*}[t]
\small
\centering

\resizebox{\textwidth}{!}{
\begin{tabular}{l|l|l|l|l|l|l|l}
\hline
\textbf{Dataset Name} & \textbf{Langs.} & \textbf{Domain} & \textbf{Topics} & \textbf{Date of Data} & \textbf{Engagements} & \textbf{Size} & \textbf{Labels} \\
\hline
LIAR \cite{Wang:2017} & En & Statements & MISC & 2007-2016 & None & 12.8k & Annotated \\
FakeNewsNet \cite{Shu:2020} & En & News, tweets & MISC & n/a & None & 23.1k, 1.9m & Query \\
CoAID \cite{Cui:2020} & En & News, tweets & C19 & 2019-2020 & Reply & 4.2k, 160k & Query \\
COVIDLies \cite{Hossain:2020} & En & Tweets & C19 & 2020 & None & 6.7k & Annotated \\
CMU-MisCOV19 \cite{Memon:2020} & En & Tweets & C19 & 2020 & None & 4.5k & Annotated \\
MM-COVID \cite{Li:2020} & 6 langs. & Tweets & C19 & n/a & Reply, retweet & 105.3k & Query \\
VaccineLies \cite{Weinzierl:2022} & En & Tweets & C19, HPV & 2019-2021 & None & 14.6k & Annotated\\
MuMin \cite{Nielsen:2022} & 41 langs. & Tweets & MISC & n/a & Reply, retweet & 21.5m & Query \\
MR2 \cite{Hu:2023} & En, Zh & Tweets, Weibo & MISC & 2017-2022 & Reply, retweet & 14.7k & Annotated \\
MiDe22 (this study) & En, Tr & Tweets & RUW, C19, IMM, MISC & 2020-2022 & Reply, retweet, like, quote & 10.3k & Annotated \\
\hline
\end{tabular}
}
\caption{\textbf{Related misinformation studies}. RUW stands for Russia-Ukraine War, C19 for COVID-19, IMM for Immigration and Refugees, HPV for Human Papilloma Virus, and MISC for Miscellaneous. The last column shows if tweets are annotated by humans, or labeled by the output of queries to Twitter API. Size is given in terms of number of tweets.}
\label{tab:other_datasets}
\end{table*}

\section{Related Work}
In this section, we provide a brief review of the existing literature and explore the methods used for the analysis and detection of misinformation, the available datasets for research purposes, and the various interventions implemented to combat the spread of misinformation.

\subsection{Misinformation Analysis}
Misinformation analysis is the process of identifying, evaluating, and understanding the spread and impact of false, misleading, or inaccurate information. Misinformation modeling covers temporal and patterns of information diffusion to analyze spread \cite{Shin:2018,Rosenfeld:2020}, and also analysis of misinformation spreads during important events such as the 2016 U.S. Election \cite{Grinberg:2019}, the COVID-19 Pandemic \cite{Ferrara:2020}, and the 2020 BLM Movement \cite{Toraman:2022b}. 

\subsection{Misinformation Detection}
Misinformation detection is a challenging task when the dynamics subject to misinformation spread are considered. The task is also studied as fake news detection \cite{Zhou:2020}, rumor detection \cite{Zubiaga:2018}, and fact/claim verification \cite{Bekoulis:2021, Guo:2022}. 

There are two important aspects of misinformation detection. First, the task mostly depends on supervised learning with a labeled dataset. 
Second, existing studies rely on different feature types for automated misinformation detection \cite{Wu:2016}. Text contents are represented in a vector or embedding space by natural language processing \cite{Oshikawa:2020} and the task is formulated as classification or regression mostly solved by deep learning models \cite{Rafiqul:2020b}. The features extracted from user profiles can be used to detect the spreaders \cite{Lee:2011}. Besides contents, there are efforts to extract features from the network structure such as network diffusion models \cite{Kwon:2014,Shu:2019a} and graph neural networks \cite{Mehta:2022}. Lastly, external knowledge sources \cite{Shi:2016, Toraman:2022} and the social context among publishers, news, and users \cite{Shu:2019b} can be integrated to the learning phase.

Rather than identifying the content with misinformation, there are efforts to detect the user accounts that would spread undesirable content such as spamming and misinformation. Social honeypot \cite{Lee:2011} is a method to identify such users by attracting them to engage with a fake account, called honeypot. There are also bots producing computer-generated content to promote misinformation \cite{Himelein:2021}. 

\subsection{Misinformation Datasets}

There are several efforts in the literature to construct a dataset for misinformation detection. The LIAR dataset \cite{Wang:2017} includes short statements from different backgrounds, annotated by PolitiFact API. News and related tweets for fact-checked events are composed in a dataset in \cite{Shu:2020}. Recently, global events and their repercussions in social media lead to the emergence of new misinformation datasets. For instance, \citet{Memon:2020} annotate tweets according to misinformation categories such as fake treatments for COVID-19. In \cite{Li:2020}, news sources are investigated for fake news in different languages. \citet{Hossain:2020} retrieve common misconceptions about COVID-19, and label tweets according to their stances against misconceptions. \citet{Weinzierl:2022} compose the vaccine version of the same dataset. Other datasets include COVID-19 healthcare misinformation \cite{Cui:2020}, and large-scale multimodal misinformation \cite{Nielsen:2022}. \cite{Hu:2023} curate annotated multimodal social media dataset for two widely-spoken languages (English and Chinese), providing reply and retweet engagements. Lastly, there are very limited datasets for low-resource languages \cite{HossainZobaer:2020, Lucas:2022} but do not exist for Turkish.

\begin{table*}[t]
    \small
    \centering
    \resizebox{\textwidth}{!}{
    \begin{tabular}{l|l}
    \textbf{Classes} & \textbf{Sample Sentence}\\
    \hline
    \textcolor{green}{True} & No, WHO's Director-General Didn't Say COVID Vaccines Are `Being Used To Kill Children'.\\
    \textcolor{red}{False} & The director-general of the WHO and I quote: ``countries are using the vaccine to kill children''.\\
    \textcolor{blue}{Other} & Africa moving toward control of COVID-19: WHO director.\\
    \hline
    \end{tabular}
    }
    \caption{Sample sentences from \texttt{MiDe22}. The event number is EN18.}
    \label{tab:sample_sentences}
\end{table*}

\subsection{Misinformation Intervention and Generative AI}
Misinformation intervention is the task of reducing the negative effects of spread in advance. One way to fight against misinformation is to spread true information by cascade modeling \cite{Budak:2011}. Other methods include detecting credible information \cite{Morstatter:2014}, cost-aware intervention \cite{Thirumuruganathan:2021}, and crowdsourcing \cite{Twitter:2022b}. However, with the recent success of transformer-based generative models, such as ChatGPT\footnote{\href{https://chat.openai.com}{https://chat.openai.com}}, it becomes more difficult for a human reader to assess and interfere with the credibility of the news source \cite{Hsu:2023}. Recent studies \cite{Zellers:2020, Giovanni:2023} reveal that social media users cannot distinguish manipulative contents generated by Generative AI \cite{Brown:2020} and humans.

\subsection{Our Differences}
In Table \ref{tab:other_datasets}, we summarize notable datasets in the literature and compare them with our dataset. We aim to provide a resource for misinformation detection and analysis, rather than intervention. Different from existing works, our study covers several recent events for misinformation analysis, including the 2022 Russia-Ukraine War, providing human-annotated tweets and user engagements on Twitter.

\begin{table*}[ht]
    \small
    \centering
    \renewcommand{\arraystretch}{1.1}
    \resizebox{\textwidth}{!}{
    \begin{tabular}{ll|rrr|rrr}
    \hline
    \multicolumn{2}{c|}{\textbf{Statistics}} & \multicolumn{3}{c|}{\textbf{\texttt{MiDe22-EN}}} & \multicolumn{3}{c}{\textbf{\texttt{MiDe22-TR}}}\\
    & & True & False & Other & True & False & Other \\ 
    \hline
    Tweets & & 727 & 1,729 & 2,828 & 669 & 1,732 & 2,663\\ 
    & Like & 11,662 & 8,587 & 33,086 & 16,594 & 24,076 & 30,446 \\
    User & Reply & 853 & 1,065 & 3,291 & 1,316 & 1,528 & 2,677 \\
    Engagements & Retweet & 2,839 & 3,127 & 9,106 & 3,055 & 5,333 & 6,442 \\
    & Quote & 339 & 451 & 2,673 & 682 & 858 & 1,649 \\
    \hline
    & Like & 16.04$\pm$168.61 & 4.97$\pm$36.65 & 11.70$\pm$153.55 & 24.80$\pm$122.33 & 13.90$\pm$85.52 & 11.43$\pm$64.16 \\
    & Reply & 1.17$\pm$11.95 & 0.62$\pm$3.82 & 1.16$\pm$11.51 & 1.97$\pm$11.66 & 0.88$\pm$4.81 & 1.01$\pm$17.50 \\
    & Retweet & 3.91$\pm$43.35 & 1.81$\pm$15.69 & 3.22$\pm$39.11 & 4.57$\pm$27.14 & 3.08$\pm$18.96 & 2.42$\pm$16.26 \\
    Average & Quote & 0.47$\pm$4.55 & 0.26$\pm$1.56 & 0.95$\pm$23.33 & 1.02$\pm$5.62 & 0.50$\pm$5.64 & 0.62$\pm$11.00 \\
    & Image & 0.10$\pm$0.30 & 0.08$\pm$0.27 & 0.13$\pm$0.34 & 0.17$\pm$0.37 & 0.22$\pm$0.42 & 0.17$\pm$0.37 \\
    & Video & 0.01$\pm$0.08 & 0.03$\pm$0.17 & 0.03$\pm$0.18 & 0.07$\pm$0.25 & 0.05$\pm$0.22 & 0.05$\pm$0.22 \\
    \hline
    \end{tabular}
    }
    \caption{The main statistics of our dataset for English (EN) and Turkish (TR). The mean and standard deviation among tweets for each attribute are given.} 
    \label{tab:data_stats}
\end{table*}

\section{Dataset Construction}\label{sec:data}

\subsection{Data Crawling}\label{sec:data_crawling}
There are 40 events under four topics per language (English and Turkish). We manually browsed fact-checking platforms (PolitiFact.com, EuVsDisinfo.eu, UsaToday.com/FactCheck for English, and Teyit.org for Turkish), and manually selected all events related to our topics at the beginning of April 2022\footnote{Some of the events were later filtered out due to the insufficient number of tweets.}. The events range from September 10th, 2020 to March 21st, 2022 in English, and October 5th, 2020 to March 11th, 2022 in Turkish. To find relevant tweets for events, we used a predetermined set of keywords for each event. At this point, we emphasize that the main criteria of keyword selection is to reach the critical mass in terms of the number of tweets for a given event. Therefore, there is not any bias stemming from keywords towards True and False labels. The details of tweet crawling and query structure are given in Appendix \ref{sec:appendix_crawling}. We collected tweets via Twitter API's Academic Research Access\footnote{Note that legacy Twitter API accesses (e.g., Academic Research Access) were deprecated in April 2023. New API accesses are provided on \href{https://developer.twitter.com/en/products/twitter-api}{https://developer.twitter.com/en/products/twitter-api}}. 

Each event is represented by 11 attributes: Event's language, id, topic, title, URL for evidence, the keywords for querying tweets, the date when evidence is provided, the start date of querying tweets, the end date of querying tweets, the keywords used while querying tweets for the Other class, sample tweet ID(s) in this event. The tweets range from September 19th, 2020 to April 5th, 2022 in English, and September 15th, 2020 to April 5th, 2022 in Turkish.  

We excluded retweets to avoid duplicates. We used Dice similarity \cite{Schutze:2008}, and applied a similarity threshold (85\%) between a newly collected tweet and previous tweets. If it exceeded the threshold, then we skipped that tweet and collected another tweet. We did not set a limit on tweet length, since misinformation can be spread by a few or no words using media objects. We kept the original contents, and provided links to the images and external URLs in tweets. We also collected all user engagements returned to our queries in the types of like, reply, retweet and quote.

\subsection{Data Annotation and Statistical Authentication}\label{sec:data_annotation}
Each tweet in the dataset is labeled according to three classes: True information, False information, and Other. The True class includes tweets with the correct information regarding the corresponding event. The False class includes tweets with misinformation on the corresponding event. The Other class includes tweets that cannot be categorized under false and true information. In general, these tweets include opinions or information related to the events, which cannot be directly judged as True or False. Sample sentences for each class label are given for the EN18 event in Table \ref{tab:sample_sentences}. The event is about the speech of T. A. Ghebreyesus, the Director-General of World Health Organization (WHO), during the opening of the WHO Academy in Lyon. Ghebreyesus stumbled over his words, first mispronouncing the word ``children'' led some people to claim he said ``kill children''. 

We assigned five annotators who were computer engineering undergraduate students. We developed an annotation tool based on INCEpTION \cite{Klie:2018} for ease of labeling. Before the annotation process started, all annotators were given a tutorial about the task. Explicit definitions of True and False tweets were provided along with the corresponding examples. We try to mitigate any bias such as political leanings and beliefs during annotation tutorials.   Annotation guidelines are detailed in the online repository\textsuperscript{\ref{note1}}. The annotations are designed so that each tweet is labeled by two annotators, and tweets are distributed randomly and almost evenly among annotators. If there was no agreement between two annotators, then a different annotator was assigned to label. We applied the majority voting in that case. If there was still no agreement among three annotators, then we removed it from the dataset by assuming that the tweet was problematic. 

Since each tweet is annotated by at least two annotators, we calculate Krippendorf’s alpha-reliability \cite{Krippendorff:1970} to measure inter-annotator agreement (IAA). The resulting alpha coefficients are 0.785 and 0.791 in English and Turkish, respectively. Regarding the study of \citet{Landis:1977}, our dataset has substantial agreement among annotators. Furthermore, the IAA scores of \texttt{MiDe22} are higher than or similar to those of existing datasets \cite{Nakamura:2019, Perez:2017, Wang:2017, Nguyen:2020}. 

\begin{figure}[t]
\centering
\subfloat[English \label{fig:wordcloud_en}]{\includegraphics[scale=0.1]{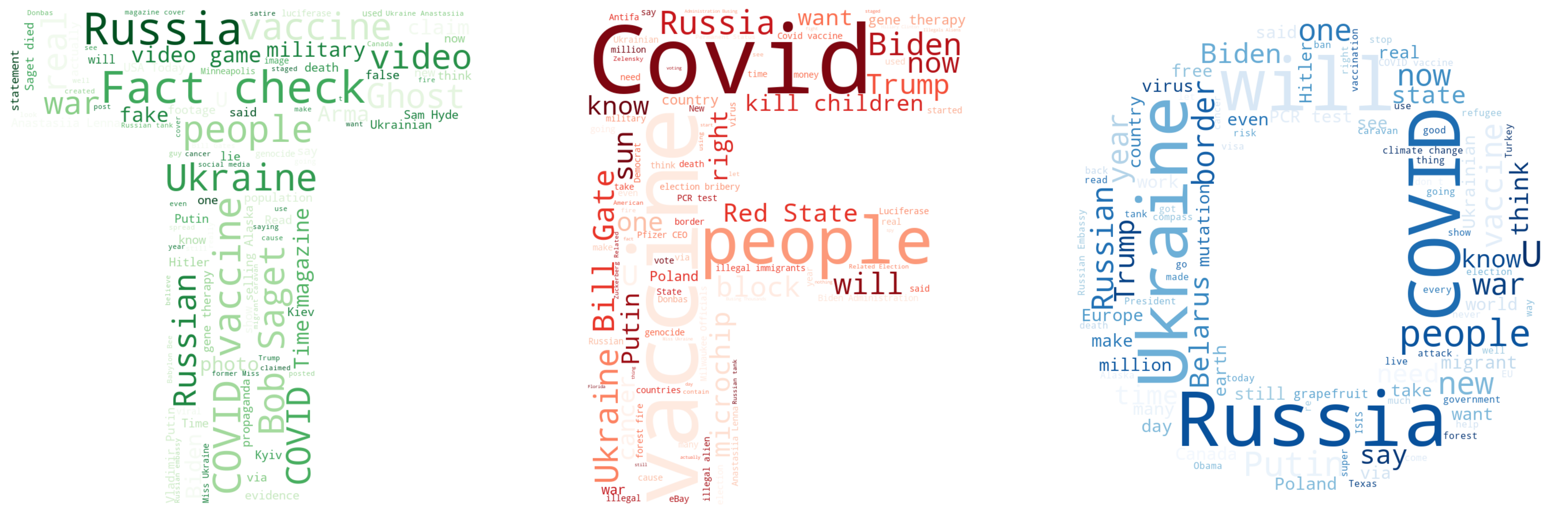}} \hspace{2.5em}
\subfloat[Turkish \label{fig:wordcloud_tr}]{\includegraphics[scale=0.1]{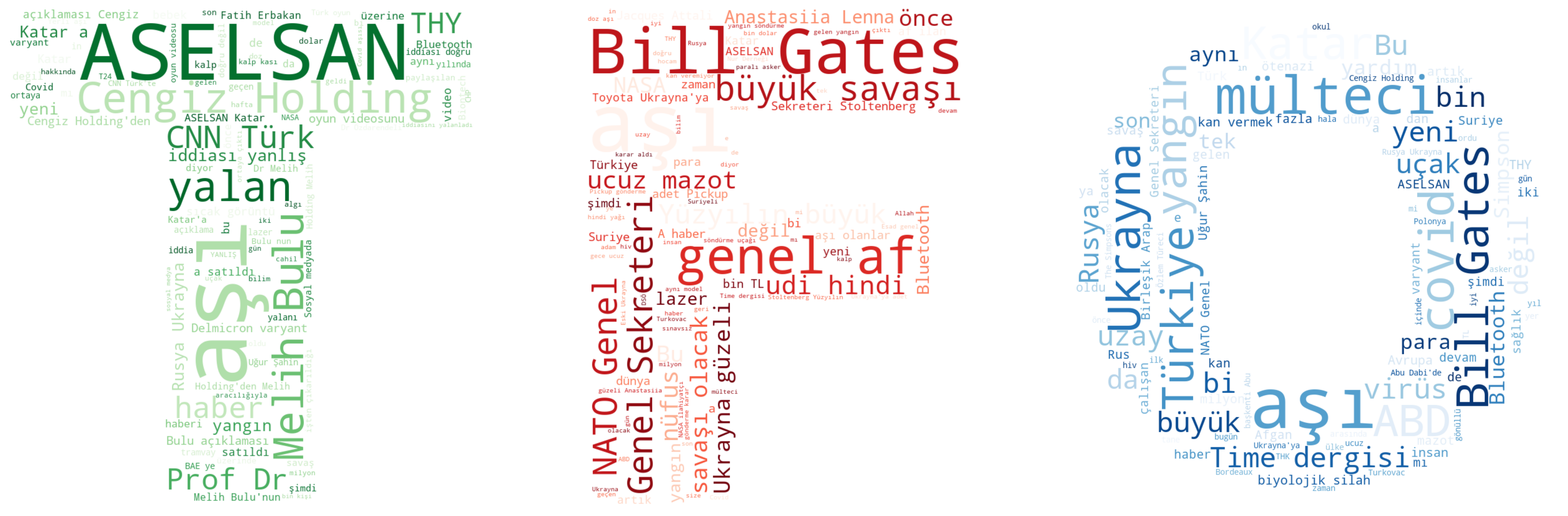}} 
\caption{Word clouds for most frequently observed keywords in the (a) English and (b) Turkish datasets for True, False, and Other. Collocations are calculated within a window size of two consecutive words.}
\end{figure}

\section{Data Analysis}

\subsection{Quantitative Analysis}
Our dataset is available in English and Turkish. The tweet counts together with the average numbers of user engagements (like, reply, retweet, and quote) are listed in Table \ref{tab:data_stats}. The average number of all types of user engagements per true tweet is higher than false tweets in both languages. On the other hand, there is no significant difference in the average number of images and videos. Nevertheless, false tweets in Turkish have more images, while false tweets in English have slightly more videos, compared to true ones.

\begin{table*}[t]
    \small
    \centering
    \renewcommand{\arraystretch}{1.1}
    \resizebox{\textwidth}{!}{
    \begin{tabular}{ll|lllll}
    & \textbf{Classes} & \multicolumn{5}{c}{\textbf{Emoji and Hashtags with \textcolor{gray}{the ratio of frequencies}}}\\
    \hline
    \multirow{6}{*}{\rotatebox[origin=c]{90}{\texttt{MiDe22-EN}}} & True & \emojiA \textcolor{gray}{0.37} & \emojiB \textcolor{gray}{0.10} & \emojiC \textcolor{gray}{0.09} & \emojiD \textcolor{gray}{0.09} & \emojiE \textcolor{gray}{0.07}\\
    & False & \emojiA \textcolor{gray}{0.24} & \emojiE \textcolor{gray}{0.15} & \emojiF \textcolor{gray}{0.11} & \emojiD \textcolor{gray}{0.08} & \emojiL \textcolor{gray}{0.08}\\
    & Other & \emojiA \textcolor{gray}{0.07} & \emojiD \textcolor{gray}{0.07} & \emojiK \textcolor{gray}{0.05} & \emojiF \textcolor{gray}{0.05} & \emojiM \textcolor{gray}{0.04}\\
    \cline{2-7}
    & True & \#Ukraine \textcolor{gray}{0.16} & \#FactCheck \textcolor{gray}{0.10}  & \#Russia \textcolor{gray}{0.09} & \#FakeNewsAlert \textcolor{gray}{0.07} & \#FactsMatter \textcolor{gray}{0.06}\\
    & False & \#Ukraine \textcolor{gray}{0.12} & \#Poland \textcolor{gray}{0.06} & \#Zelensky \textcolor{gray}{0.05} & \#COVID19 \textcolor{gray}{0.05} & \#Russia \textcolor{gray}{0.04}\\
    & Other & \#Ukraine \textcolor{gray}{0.15} & \#Poland \textcolor{gray}{0.06} & \#Zelensky \textcolor{gray}{0.05} & \#COVID19 \textcolor{gray}{0.05} & \#Russia \textcolor{gray}{0.04}\\
    
    \hline \hline
    
    \multirow{6}{*}{\rotatebox[origin=c]{90}{\texttt{MiDe22-TR}}} & True & \emojiG \textcolor{gray}{0.41} & \emojiA \textcolor{gray}{0.15} & \emojiJ \textcolor{gray}{0.14} & \emojiN \textcolor{gray}{0.09} & \emojiH \textcolor{gray}{0.08}\\
    & False & \emojiA \textcolor{gray}{0.18} & \emojiE \textcolor{gray}{0.10} & \emojiD \textcolor{gray}{0.08} & \emojiO \textcolor{gray}{0.07} & \emojiF \textcolor{gray}{0.06}\\
    & Other & \emojiA \textcolor{gray}{0.17} & \emojiO \textcolor{gray}{0.10} & \emojiE \textcolor{gray}{0.08} & \emojiD \textcolor{gray}{0.08} & \emojiK \textcolor{gray}{0.06}\\
    \cline{2-7}
    & True & \#Ukrayna \textcolor{gray}{0.10} & \#sondakika \textcolor{gray}{0.05} & \#Rusya \textcolor{gray}{0.05} & \#cnnturk \textcolor{gray}{0.05} & \#haber \textcolor{gray}{0.04} \\
    & False & \#Ukrayna \textcolor{gray}{0.06} & \#UkraineRussiaWar \textcolor{gray}{0.06} & \#SONDAKİKA \textcolor{gray}{0.04} & \#Ukraine \textcolor{gray}{0.03} & \#worldwar3 \textcolor{gray}{0.03}\\
    & Other & \#Ukrayna \textcolor{gray}{0.06} & \#Rusya \textcolor{gray}{0.05} & \#uzay \textcolor{gray}{0.03} & \#mülteci \textcolor{gray}{0.03} & \#SONDAKİKA \textcolor{gray}{0.03}\\ \hline
    \end{tabular}
    }
    \caption{\textbf{Top-5 most frequent emoji and hashtags for each class (row) with their frequency ratios.} The ratio is the number of emoji/hashtag divided by the number of tweets with that emoji/hashtag.}
    \label{tab:top_emoji_and_url}
\end{table*}

\subsection{Content Analysis}
Figure \ref{fig:wordcloud_en} and \ref{fig:wordcloud_tr} show the word clouds for each class (true, false, and other). Although the dataset includes several topics, COVID-related keywords are observed more in false tweets for both languages (e.g. ``aşı'' translated as ``vaccine''), and also political figures (e.g. Biden, Trump, and Putin). On the contrary, we observe fact-checking keywords in true tweets (e.g. ``yalan'' translated as ``lie'').

We also provide the top five most frequently observed emojis and hashtags in Table \ref{tab:top_emoji_and_url}. When smiling and laughing emojis are discarded, true tweets include mostly cross signs that would represent false information. False tweets contain the pointing down emoji that would point a message to readers, and also a thinking emoji that would emphasize the false information to readers. In terms of hashtags, we find that most of the hashtags are related to the 2022 Russia-Ukraine War. In English, there are fact-checking hashtags in true tweets (e.g. \#FakeNewsAlert), while a similar kind of hashtag is not observed in Turkish.

\begin{figure*}[t]
\centering
\subfloat[Russia-Ukraine War in English \label{fig:temp_en_ukr}]{\includegraphics[width=.68\columnwidth]{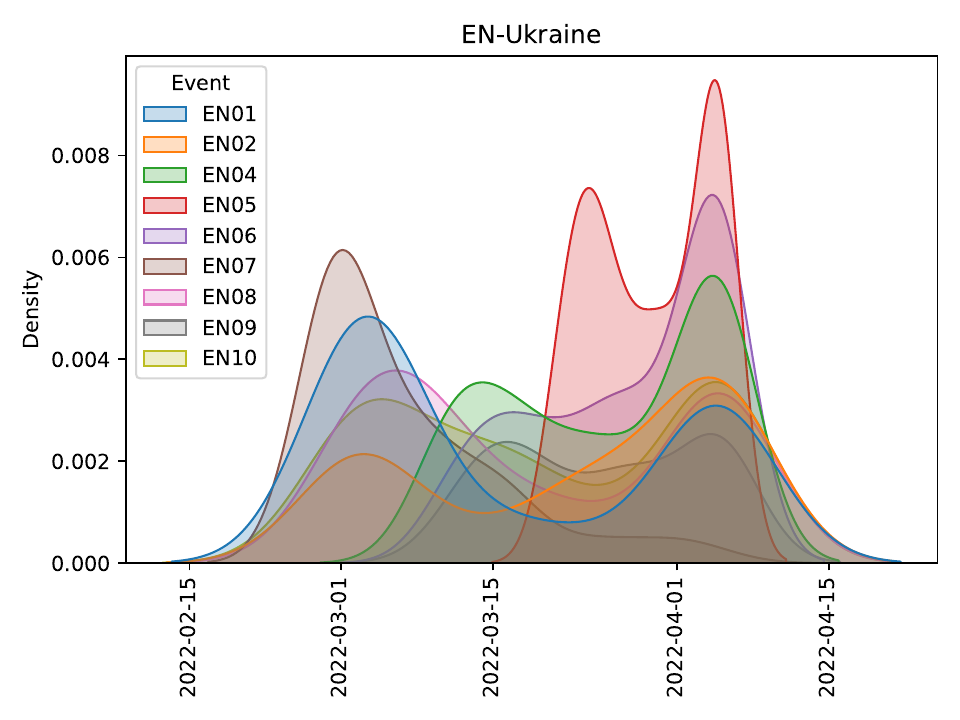}} 
\subfloat[COVID-19 in English \label{fig:temp_en_covid}]{\includegraphics[width=.68\columnwidth]{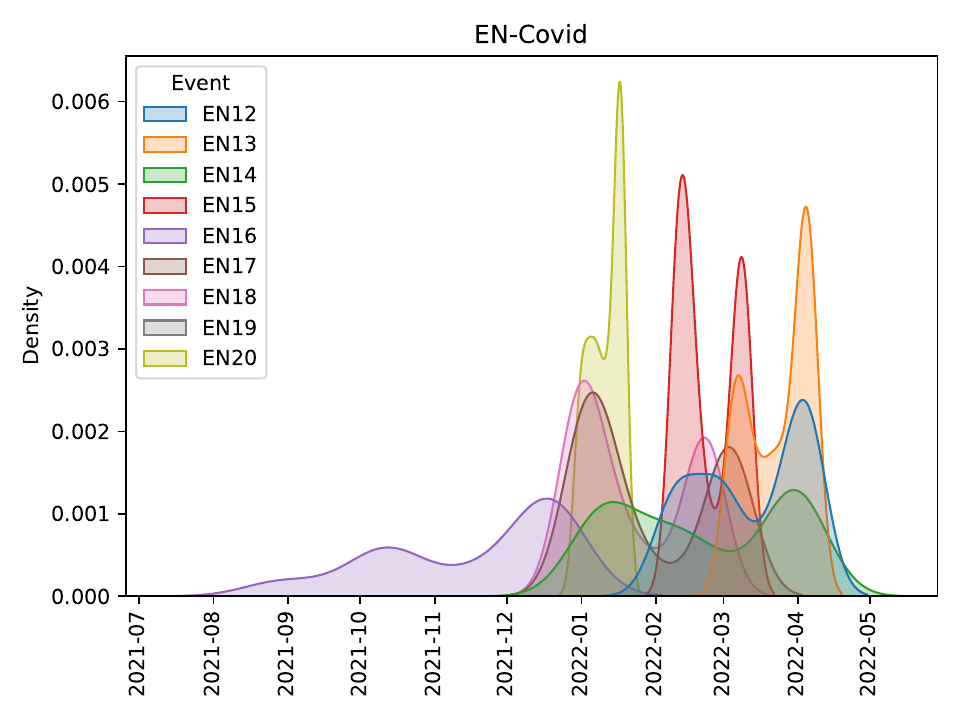}}\hfil
\subfloat[Immigration in English \label{fig:temp_en_immig}]{\includegraphics[width=.68\columnwidth]{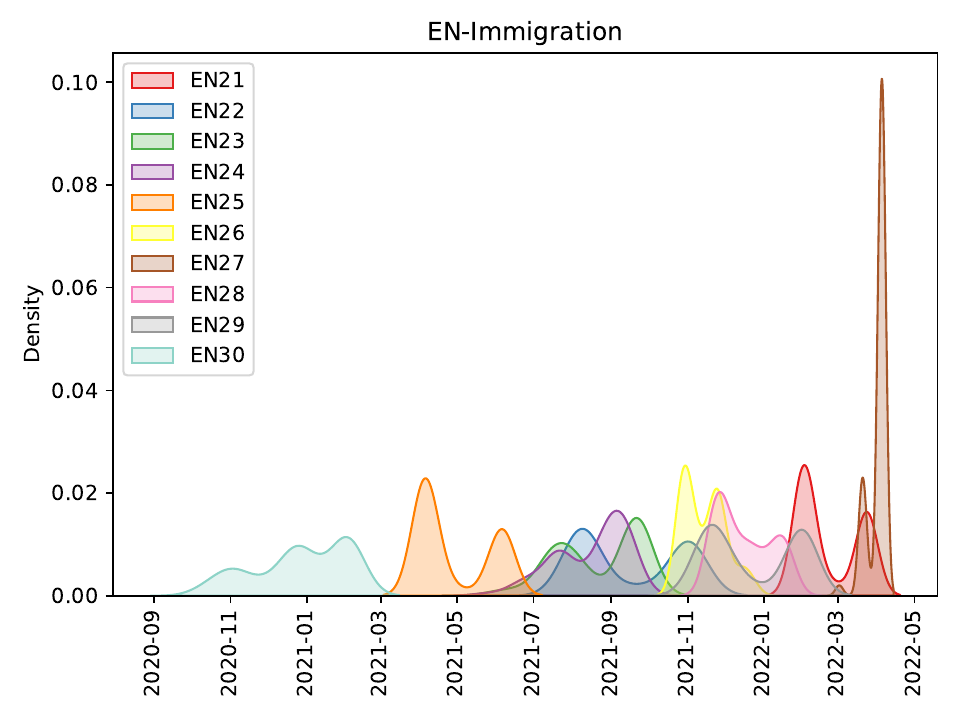}}\hfil 

\subfloat[Russia-Ukraine War in Turkish
\label{fig:temp_tr_ukr}]{\includegraphics[width=.68\columnwidth]{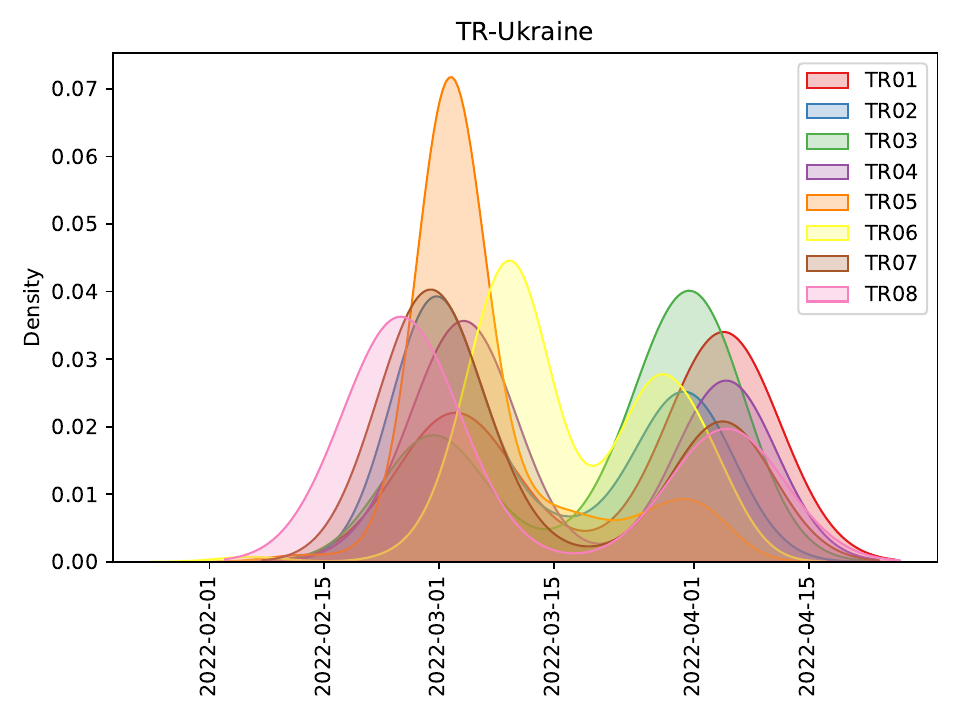}}
\subfloat[COVID-19 in Turkish \label{fig:temp_tr_covid}]{\includegraphics[width=.68\columnwidth]{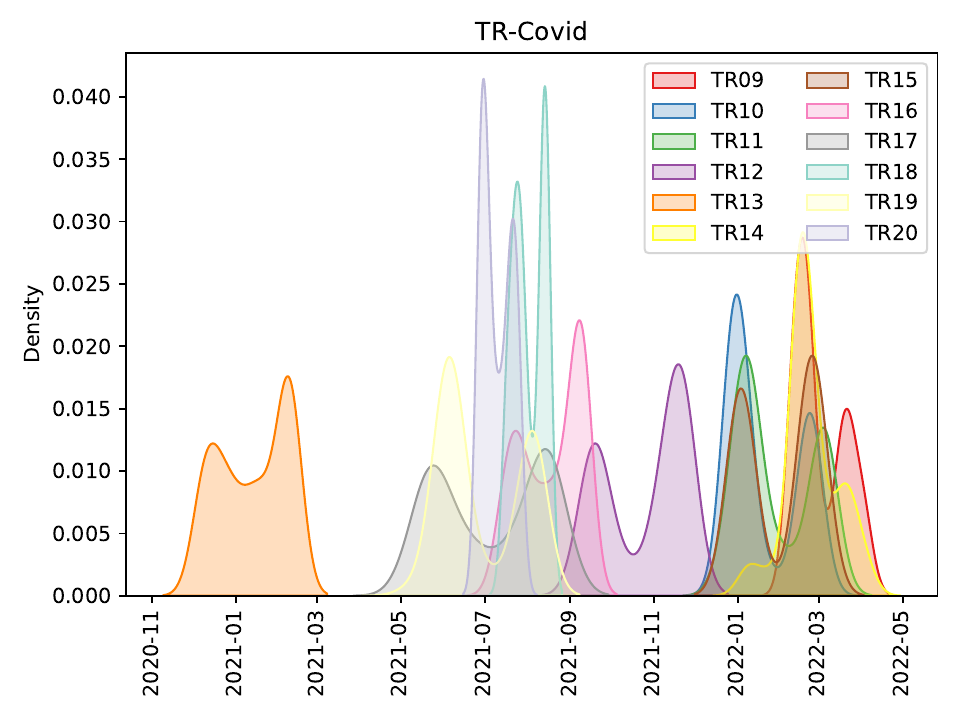}}\hfil   
\subfloat[Immigration in Turkish \label{fig:temp_tr_immig}]{\includegraphics[width=.68\columnwidth]{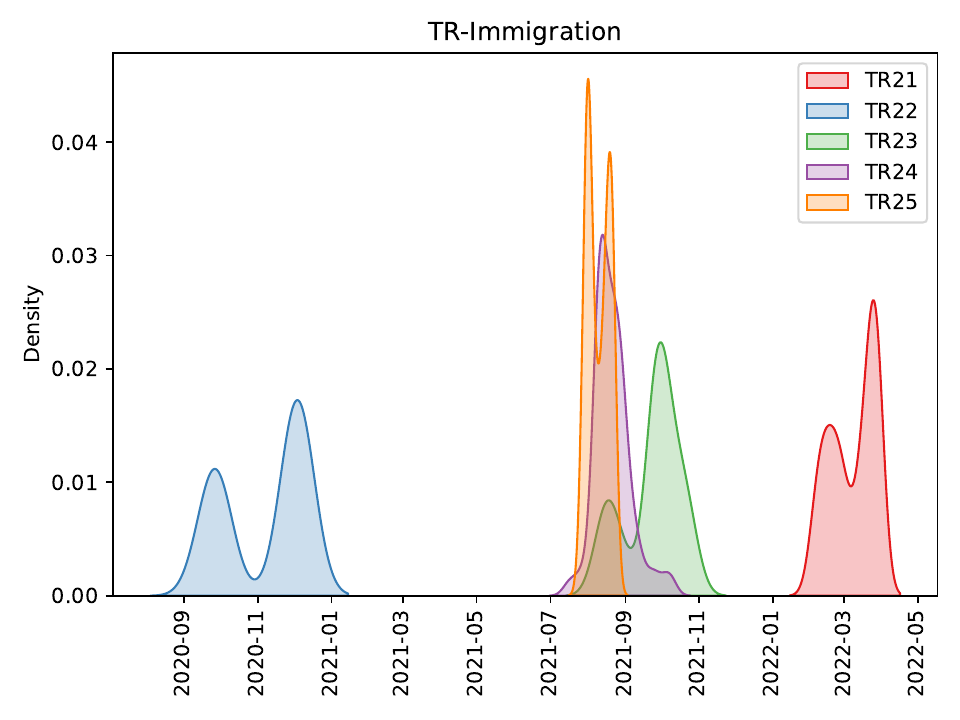}}\hfil
\caption{\textbf{Temporal distribution of tweets by topics}. The y-axis represents the density of tweet counts. The x-axis represents the date that tweets are shared. The events EN03 and EN11 are neglected due to the shorter time range. }\label{fig:all_temporal}
\end{figure*}

\subsection{Temporal Analysis}

We provide the distribution of the tweets for each topic in Figure \ref{fig:all_temporal}. Most of the events gain popularity rapidly, reach their peak, and fall from the grace after a while. Most of the distributions exhibit a bimodal Gaussian shape, meaning that there is a second peak point (local or global) for the distribution. There is a stimulus that makes the event regain its popularity. In our early analyses, we find that the dates of the turning point in distributions coincide with the dates of fact-checking news. 

When we examine the distribution of events according to tweet posting date, we observe both similar and different patterns among topics. For instance, COVID-19 events can last up to six months (Figure \ref{fig:temp_en_covid}), since it is a long-term incident. A similar pattern also exists in Turkish with events lasting more than four months (Figure \ref{fig:temp_tr_covid}).
The Russia-Ukraine War is a fresh topic, covering a relatively shorter time period. However, Figures \ref{fig:temp_en_ukr} and \ref{fig:temp_tr_ukr} show that there are different events that could lead to the spread of misinformation in this short period of time. Overall, we argue that detection algorithms can be developed based on the event's life span, e.g. user engagements for the long-term and context for short-term events.

\section{Experiments}
In order to understand if the constructed dataset is adequate in terms of task difficulty, we target misinformation detection using only tweet text. We leave utilization of user engagements for future work. We implement total of eight benchmark models (see Table \ref{tab:first_TurkishResults}) from the following model families:

\noindent \textbf{Bag-of-Words}: We consider conventional machine learning classifiers based on the bag-of-words model since tweets can include specific terms and phrases used for reporting manipulated news (e.g. "Did you know?") and correcting falsehood (e.g. "FactCheck"). We implement a linear Support Vector Machine (SVM) \cite{Vapnik:1999} with TD-IDF vectors of each tweet using scikit-learn \cite{Scikit:Learn}. SVM is trained with a stopping criterion, i.e., 1e-3 tolerance. The remaining parameters are selected default.

\noindent \textbf{Neural Models}: We implement Long Short-term Memory (LSTM) \cite{Hochreiter:1996} and Bi-directional Long Short-Term Memory (BiLSTM) \cite{Graves:2005} with PyTorch \cite{Paszke:2019}. The embedding size is 125, and there are 50 units in each layer. After the LSTM layers, there is a dense layer with perceptrons of the same size of units. Next, there is a dropout layer with a probability of 0.5. We use the sigmoid activation function. They are trained during 20 epochs, where we set a learning rate of 1e-3 with a batch size of 16.

\noindent \textbf{Transformer-based Language Models}: We use BERT base uncased \cite{Devlin:2019} and DeBERTa \cite{He:2021} pretrained with English corpus, BERTurk uncased base model \cite{Schweter:2020} for Turkish corpus, and mBERT base uncased \cite{Devlin:2019} and XLM-R base \cite{Conneau:2020} for multilingual corpus, by HuggingFace \cite{Wolf:2020}. We use cross-entropy loss and set a learning rate of 5e-5 with a batch size of 16 during 10 epochs. The number of tokens is set to 128 with padding and truncation, where each tweet is an input sequence.

We apply stratified (in terms of both classes and events) 5-fold cross-validation to get an average performance score for robustness. The training of SVM, LSTM, and BiLSTM is performed on Intel Core i9-10900X 3.70GHz CPU with 20 cores with 128 GB memory. The fine-tuning of large language models is performed on a single NVIDIA RTX A4000.

\begin{table*}[t]
    \small
    \centering
    \resizebox{\textwidth}{!}{
    \begin{tabular}{l|lll|lll}
    \hline
    \multirow{ 2}{*}{\textbf{Model}} & \multicolumn{3}{c|}{\textbf{\texttt{MiDe22-EN}}}& \multicolumn{3}{c}{\textbf{\texttt{MiDe22-TR}}}\\
    
     & Precision & Recall & F1 & Precision & Recall & F1 \\
    \hline
    
    SVM & 79.29$\pm$0.9 & 79.00$\pm$0.8 & 78.33$\pm$0.9 & 78.78$\pm$1.1 & 78.60$\pm$1.1 & 78.20$\pm$1.1 \\
    
    LSTM & 72.71$\pm$2.2 & 72.94$\pm$2.0 & 72.24$\pm$1.8 & 72.59$\pm$0.9 & 72.71$\pm$0.9 & 72.23$\pm$0.8 \\
    
    BiLSTM & 73.31$\pm$0.7 & 73.61$\pm$0.8 & 73.30$\pm$0.7 & 74.16$\pm$0.7 & 74.21$\pm$0.8 & 73.77$\pm$0.9\\
    
    BERT & 82.44$\pm$0.8 & 82.33$\pm$1.0 & 82.35$\pm$0.9 & 78.63$\pm$1.3 & 78.39$\pm$1.4 & 78.46$\pm$1.4 \\
    
    DeBERTa & \textbf{84.04$\pm$0.7} & \textbf{83.94$\pm$0.7} & \textbf{83.95$\pm$0.7} & 76.03$\pm$2.2 & 75.14$\pm$2.3 & 75.30$\pm$2.2 \\
    
    mBERT & 78.87$\pm$2.9 & 78.37$\pm$4.1 & 78.21$\pm$4.3 & 78.63$\pm$1.0 & 78.20$\pm$1.2 & 78.26$\pm$1.2\\
    
    XLM-R & 79.19$\pm$3.0 & 78.94$\pm$3.0 & 79.01$\pm$3.0 & \textbf{83.18$\pm$1.1} & \textbf{82.76$\pm$1.1} & \textbf{82.82$\pm$1.1}\\
    
    BERTurk & 78.31$\pm$1.5 & 78.36$\pm$1.6 & 78.30$\pm$1.6 & 82.89$\pm$1.1 & 82.52$\pm$1.1 & 82.58$\pm$1.1 \\
    \hline
    \end{tabular}
    }
    \caption{ \textbf{The results of benchmark models for Misinformation Detection on \texttt{MiDe22}}. The average score of five folds with standard deviation is reported in terms of weighted precision, recall, and F1 scores. The best scores for each dataset and metric are given in bold.}
    \label{tab:first_TurkishResults}
\end{table*}

\subsection{Experimental Results}
We report the performance of models in Table \ref{tab:first_TurkishResults}. We observe that Transformer-based language models outperform conventional methods (SVM, LSTM, and BiLSTM) in both languages. However, SVM has a better performance than LSTM and BiLSTM. The reason could be the distribution of words in the false and true tweets. SVM is trained on a Bag-of-Words model where features represent the importance score of individual words. 

Among language models, DeBERTa has the highest performance in English, while a multilingual model, XLM-R, in Turkish. This observation can show the generalization capability of multilingual models for low-resource languages, such as Turkish, in misinformation detection. 

The performance of misinformation detection in terms of F1-Score can be observed differently in other annotated datasets: 90.07 in \cite{Weinzierl:2022}, 50.20 in \cite{Hossain:2020}, and 83.95 in our study (no F1 score reported in \cite{Wang:2017} and no detection score in \cite{Memon:2020}). We argue that the performance depends on datasets since misinformation detection is a dynamic task where context changes rapidly. The context can be integrated into the learning phase via knowledge sources \cite{Pan:2018, Toraman:2022} to adapt to the dynamic nature of the misinformation task.

\section{Discussion}
\subsection{Possible Use Cases}
The dataset can be used for several tasks in natural language processing, information retrieval, and computer vision. Misinformation detection with textual features \cite{Su:2020} or visual features \cite{Cao:2020} is the primary objective of this dataset. Multimodal approaches with both textual and visual features have also promising results \cite{Khattar:2019}. Other opportunities include the analysis of information diffusion \cite{Shin:2018} and bot accounts using tweet conversations \cite{Cetinkaya:2020} since the dataset has the user engagements for all tweets. Furthermore, the efforts to detect the events of social media posts, i.e. event or topic detection \cite{Sahinuc:2021}, can benefit due to the variety of events in the dataset.

\subsection{Difficulties Encountered}
Finding relevant tweets to events was a challenging task. We run different queries with different keywords to fetch the highest number of relevant tweets. Although we used Twitter Academic Access API, we could not increase the number of tweets relevant to events. Another difficulty was the guidance of annotators in this dataset. We tried to guide the annotators to be as objective and unbiased as possible by providing a guidelines document and a dedicated live video seminar, where we explained the events, claims and evidences, annotation tool, and example annotations.

\section{Conclusion}
We curate a multi-event tweet dataset for misinformation detection that has novelty in terms of the variety of languages (English and Turkish), topics (various topics and 40 events per language), and engagements (like, reply, retweet, and quote). We further analyze the dataset and provide benchmark experiments including the performances of state-of-the-art models. We publish the dataset and the files related to the dataset curation for transparency. They provide new opportunities for researchers from different backgrounds including but not limited to natural language processing, social network analysis, and computer vision. 

In future work, we plan to develop new models on our dataset for various tasks such as multimodal detection and adversarial attacks for misinformation. Cross-lingual misinformation spread is another opportunity since our dataset covers two languages with overlapping events. We can also extend our study to other social media platforms for cross-platform misinformation detection. 

\section{Limitations}
We acknowledge a set of limitations in this study. First, creating a misinformation dataset is more difficult than other types of tweet datasets due to the regulations of social media platforms. Making the dataset balanced in terms of labels can be therefore challenging. Second, we decided on the events included in the dataset manually by browsing the fact-checking platforms such as PolitiFact.com and Teyit.org. Furthermore, human annotation is a costly and laborious process. 

In this study, we labor five annotators to label tweets due to budget and time limitations. The annotators were given careful guidelines on the topics and definitions of class labels. However, the dataset can still reflect their personal biases and interpretations to some extent. Recent advances in generative AI can be also integrated to generate label annotations \cite{Zhu:2023}. Lastly, our study focuses on the English and Turkish languages only, which might reflect the cultural biases exposed by newsletters and fact-checkers. There could be different instances for the same topics in other languages.

\begin{table*}[t]
\small
\renewcommand{\arraystretch}{1.1}
\centering
\resizebox{\textwidth}{!}{
\begin{tabular}{ll}
\hline
\textbf{Language} &  English \\
\textbf{ID} &  EN2 \\
\textbf{Topic} & Ukraine \\
\textbf{Title} & Viral clip shows 'Arma 3' video game not war between Russia and Ukraine \\
\textbf{Evidence URL} & https://www.usatoday.com/story/news/factcheck/2022/02/21/.../6879521001/ \\
\textbf{Query keywords} & arma 3 russia ukraine \\
\textbf{Evidence date} & 2022-02-21 \\
\textbf{Query start date} & 2021-12-21 \\ 
\textbf{Query end date} & 2022-04-06 \\
\textbf{Other keywords} & russia ukraine war video \\
\textbf{Sample tweet(s)} & 1499460925253832707 \\
\hline
\textbf{Query-1} & arma 3 russia ukraine lang:en (has:media OR has:geo) -is:retweet 2021-12-21 2022-04-06 100 \\
\textbf{Query-2} & arma 3 russia ukraine lang:en -is:retweet 2021-12-21 2022-04-06 100 \\
\textbf{Query-3 for Other} & russia ukraine war video lang:en (has:media OR has:geo) -is:retweet 2021-12-21 2022-04-06  50 \\
\textbf{Query-4 for Other} & russia ukraine war video lang:en -is:retweet 2021-12-21 2022-04-06 50\\
\hline
\end{tabular}
}
\caption{An example event in the dataset. A part of URL is cropped due to space constraints.}
\label{tab:sample_event}
\end{table*}

\section{Ethical Concerns}
We consider the ethical concerns regarding the stakeholders in misinformation detection \cite{Neumann:2022}. First, all sources of information (tweet author) should be treated equally. We collected the tweets returned to our API queries without discriminating or selecting authors. Second, subjects of information (the subject in tweet content) should be represented fairly and accurately. Since tweets may include false claims about the subject of information, we included true tweets that refute the claims as well. Third, all seekers of information (the audience of tweet authors) should obtain relevant and high-quality information. Distributive justice is out of context since our focus is to detect misinformation not to distribute tweets to the audience. Lastly, individuals/organizations should generate fair evidence with testimonial justice. We assigned annotators to label the data according to a guidelines document that includes the details of events, claims, and corrections with sources of evidence\footnote{We relied on trust-worthy fact-checking platforms as sources of evidence: https://www.politifact.com, https://euvsdisinfo.eu, and https://eu.usatoday.com/news/factcheck for English, and https://teyit.org for Turkish.}.

We obtain an internal IRB approval for our misinformation detection dataset study, which includes the approval of two reviewers.

In order to provide transparency \cite{Bender:2021, Baeza-Yates:2022}, we publish the files related to data crawling and annotation: The queries and details of the events, the annotation guidelines document, video seminar recording for directing annotators, and the details of the annotation tool\textsuperscript{\ref{note1}}.

\appendix
\section{Appendix}
\label{sec:appendix}

\subsection{Tweet Crawling}
\label{sec:appendix_crawling}
An example query for crawling tweets from Twitter API for a specific event is given in Table \ref{tab:sample_event}. We manually determined the events and query keywords by browsing events in fact checking web pages. The motivation for using a different keyword set for the Other class is that we might not find irrelevant tweets or tweets with no information with the query keywords prepared for the True and False classes. The start and end dates of querying tweets are selected before and after two months of the evidence date, unless restricted by the crawling date (2022-04-06).

We run four consecutive queries for each event. We first collected tweets with media object (image, video, or GIF) and geographic location tags (Query-1). If the number of such tweets was not enough to fulfill the number of target tweets (50 tweets per class, total of 100 tweets for the True and False classes), then we collected tweets without media objects and geographic location tags (Query-2). After running queries for the True and False classes, we collected tweets for the Other class with the same approach by first searching for media objects (Query-3) and then regular tweets (Query-4).

We set the highest number of tweets to be collected for each class (true, false, other) to 50 tweets to provide a balance among classes and limit the total number of tweets to be annotated. 

\subsection{Event List}
\label{sec:appendix_event_list}
There are four topics in the dataset. The topics are the 2022 Russia-Ukraine War, COVID-19 pandemic, Refugees (Immigration), and Miscellaneous. There are 40 events under four topics for both languages. The list of events along with their topics are published online\footref{note1}.

\subsection{User Engagements}
\label{sec:appendix_engagement_list}
The detailed list of tweets and user engagements (like, retweet, reply, and quote) per event are published online\footref{note1}.

\subsection{Annotation Tool}
\label{sec:appendix_annotation_tool}
The details of the annotation tool are published online\footref{note1}.

\section{Bibliographical References}\label{sec:reference}

\balance
\bibliographystyle{lrec-coling2024-natbib}
\bibliography{references}

\begin{thebibliography}{72}
\expandafter\ifx\csname natexlab\endcsname\relax\def\natexlab#1{#1}\fi

\bibitem[{Baeza-Yates(2022)}]{Baeza-Yates:2022}
Ricardo Baeza-Yates. 2022.
\newblock Ethical challenges in ai.
\newblock In \emph{Proceedings of the Fifteenth ACM International Conference on
  Web Search and Data Mining}, pages 1--2.

\bibitem[{Bekoulis et~al.(2021)Bekoulis, Papagiannopoulou, and
  Deligiannis}]{Bekoulis:2021}
Giannis Bekoulis, Christina Papagiannopoulou, and Nikos Deligiannis. 2021.
\newblock A review on fact extraction and verification.
\newblock \emph{ACM Computing Surveys (CSUR)}, 55(1):1--35.

\bibitem[{Bender et~al.(2021)Bender, Gebru, McMillan-Major, and
  Shmitchell}]{Bender:2021}
Emily~M. Bender, Timnit Gebru, Angelina McMillan-Major, and Shmargaret
  Shmitchell. 2021.
\newblock \href {https://doi.org/10.1145/3442188.3445922} {On the dangers of
  stochastic parrots: Can language models be too big?}
\newblock In \emph{Proceedings of the 2021 ACM Conference on Fairness,
  Accountability, and Transparency}, FAccT '21, page 610–623, New York, NY,
  USA. Association for Computing Machinery.

\bibitem[{Brown et~al.(2020)}]{Brown:2020}
Tom Brown et~al. 2020.
\newblock Language models are few-shot learners.
\newblock In \emph{Advances in Neural Information Processing Systems},
  volume~33, pages 1877--1901. Curran Associates, Inc.

\bibitem[{Budak et~al.(2011)Budak, Agrawal, and El~Abbadi}]{Budak:2011}
Ceren Budak, Divyakant Agrawal, and Amr El~Abbadi. 2011.
\newblock \href {https://doi.org/10.1145/1963405.1963499} {Limiting the spread
  of misinformation in social networks}.
\newblock In \emph{Proceedings of the 20th International Conference on World
  Wide Web}, WWW '11, page 665–674, New York, NY, USA. Association for
  Computing Machinery.

\bibitem[{Cao et~al.(2020)Cao, Qi, Sheng, Yang, Guo, and Li}]{Cao:2020}
Juan Cao, Peng Qi, Qiang Sheng, Tianyun Yang, Junbo Guo, and Jintao Li. 2020.
\newblock Exploring the role of visual content in fake news detection.
\newblock \emph{Disinformation, Misinformation, and Fake News in Social Media},
  pages 141--161.

\bibitem[{{\c{C}}etinkaya et~al.(2020){\c{C}}etinkaya, Toroslu, and
  Davulcu}]{Cetinkaya:2020}
Yusuf~M{\"u}cahit {\c{C}}etinkaya, {\.I}smail~Hakk{\i} Toroslu, and Hasan
  Davulcu. 2020.
\newblock Developing a {T}witter bot that can join a discussion using
  state-of-the-art architectures.
\newblock \emph{Social Network Analysis and Mining}, 10(1):1--21.

\bibitem[{CollinsDictionary(2017)}]{Collins:2017}
CollinsDictionary. 2017.
\newblock \href
  {https://blog.collinsdictionary.com/language-lovers/collins-2017-word-of-the-year-shortlist/}
  {Collins 2017 word of the year shortlist}.
\newblock (Accessed Oct 19, 2023).

\bibitem[{Conneau et~al.(2020)}]{Conneau:2020}
Alexis Conneau et~al. 2020.
\newblock \href {https://doi.org/10.18653/v1/2020.acl-main.747} {Unsupervised
  cross-lingual representation learning at scale}.
\newblock In \emph{Proceedings of the 58th Annual Meeting of the Association
  for Computational Linguistics, {ACL} 2020, Online, July 5-10, 2020}, pages
  8440--8451. Association for Computational Linguistics.

\bibitem[{\c{S}ahinu\c{c} and Toraman(2021)}]{Sahinuc:2021}
Furkan \c{S}ahinu\c{c} and Cagri Toraman. 2021.
\newblock \href {https://doi.org/10.1007/978-3-030-72240-1_50} {Tweet length
  matters: A comparative analysis on topic detection in microblogs}.
\newblock In \emph{Advances in Information Retrieval: 43rd European Conference
  on IR Research, ECIR 2021, Virtual Event, March 28 – April 1, 2021,
  Proceedings, Part II}, page 471–478, Berlin, Heidelberg. Springer-Verlag.

\bibitem[{Cui and Lee(2020)}]{Cui:2020}
Limeng Cui and Dongwon Lee. 2020.
\newblock Co{AID}: {COVID}-19 healthcare misinformation dataset.
\newblock \emph{arXiv preprint arXiv:2006.00885}.

\bibitem[{Devlin et~al.(2019)Devlin, Chang, Lee, and Toutanova}]{Devlin:2019}
Jacob Devlin, Ming-Wei Chang, Kenton Lee, and Kristina Toutanova. 2019.
\newblock \href {https://doi.org/10.18653/v1/N19-1423} {{BERT}: Pre-training of
  deep bidirectional transformers for language understanding}.
\newblock In \emph{Proceedings of the 2019 Conference of the North {A}merican
  Chapter of the Association for Computational Linguistics: Human Language
  Technologies, Volume 1 (Long and Short Papers)}, pages 4171--4186,
  Minneapolis, Minnesota. Association for Computational Linguistics.

\bibitem[{D’Ulizia et~al.(2021)D’Ulizia, Caschera, Ferri, and
  Grifoni}]{DUlizia:2021}
Arianna D’Ulizia, Maria~Chiara Caschera, Fernando Ferri, and Patrizia
  Grifoni. 2021.
\newblock Fake news detection: A survey of evaluation datasets.
\newblock \emph{PeerJ Computer Science}, 7:e518.

\bibitem[{Ferrara et~al.(2020)Ferrara, Cresci, and Luceri}]{Ferrara:2020}
Emilio Ferrara, Stefano Cresci, and Luca Luceri. 2020.
\newblock Misinformation, manipulation, and abuse on social media in the era of
  {COVID}-19.
\newblock \emph{Journal of Computational Social Science}, 3(2):271--277.

\bibitem[{Graves and Schmidhuber(2005)}]{Graves:2005}
Alex Graves and J{\"{u}}rgen Schmidhuber. 2005.
\newblock \href {https://doi.org/10.1016/j.neunet.2005.06.042} {Framewise
  phoneme classification with bidirectional {LSTM} and other neural network
  architectures}.
\newblock \emph{Neural Networks}, 18(5-6):602--610.

\bibitem[{Grinberg et~al.(2019)Grinberg, Joseph, Friedland, Swire-Thompson, and
  Lazer}]{Grinberg:2019}
Nir Grinberg, Kenneth Joseph, Lisa Friedland, Briony Swire-Thompson, and David
  Lazer. 2019.
\newblock \href {https://doi.org/10.1126/science.aau2706} {Fake news on
  {T}witter during the 2016 {U.S} presidential election}.
\newblock \emph{Science}, 363(6425):374--378.

\bibitem[{Guo et~al.(2022)Guo, Schlichtkrull, and Vlachos}]{Guo:2022}
Zhijiang Guo, Michael Schlichtkrull, and Andreas Vlachos. 2022.
\newblock \href {https://doi.org/10.1162/tacl_a_00454} {A survey on automated
  fact-checking}.
\newblock \emph{Transactions of the Association for Computational Linguistics},
  10:178--206.

\bibitem[{He et~al.(2021)He, Gao, and Chen}]{He:2021}
Pengcheng He, Jianfeng Gao, and Weizhu Chen. 2021.
\newblock \href {https://doi.org/10.48550/ARXIV.2111.09543} {De{BERT}a{V3}:
  Improving {D}e{BERT}a using electra-style pre-training with
  gradient-disentangled embedding sharing}.

\bibitem[{Himelein-Wachowiak et~al.(2021)}]{Himelein:2021}
McKenzie Himelein-Wachowiak et~al. 2021.
\newblock \href {https://doi.org/10.2196/26933} {Bots and misinformation spread
  on social media: Implications for {COVID}-19}.
\newblock \emph{J Med Internet Res}, 23(5):e26933.

\bibitem[{Hochreiter and Schmidhuber(1996)}]{Hochreiter:1996}
Sepp Hochreiter and J{\"{u}}rgen Schmidhuber. 1996.
\newblock {LSTM} can solve hard long time lag problems.
\newblock In \emph{Advances in Neural Information Processing Systems 9, NIPS,
  Denver, CO, USA, December 2-5, 1996}, pages 473--479. {MIT} Press.

\bibitem[{Hossain et~al.(2020{\natexlab{a}})Hossain, Rahman, Islam, and
  Kar}]{HossainZobaer:2020}
Md~Zobaer Hossain, Md~Ashraful Rahman, Md~Saiful Islam, and Sudipta Kar.
  2020{\natexlab{a}}.
\newblock \href {https://aclanthology.org/2020.lrec-1.349} {{B}an{F}ake{N}ews:
  A dataset for detecting fake news in {B}angla}.
\newblock In \emph{Proceedings of the Twelfth Language Resources and Evaluation
  Conference}, pages 2862--2871, Marseille, France. European Language Resources
  Association.

\bibitem[{Hossain et~al.(2020{\natexlab{b}})Hossain, Logan~IV, Ugarte,
  Matsubara, Young, and Singh}]{Hossain:2020}
Tamanna Hossain, Robert~L. Logan~IV, Arjuna Ugarte, Yoshitomo Matsubara, Sean
  Young, and Sameer Singh. 2020{\natexlab{b}}.
\newblock \href {https://doi.org/10.18653/v1/2020.nlpcovid19-2.11}
  {{COVIDL}ies: Detecting {COVID}-19 misinformation on social media}.
\newblock In \emph{Proceedings of the 1st Workshop on {NLP} for {COVID}-19
  (Part 2) at {EMNLP} 2020}, Online. Association for Computational Linguistics.

\bibitem[{Hsu and Thompson(2023)}]{Hsu:2023}
T.~Hsu and S.~A. Thompson. 2023.
\newblock \href
  {https://www.nytimes.com/2023/02/08/technology/ai-chatbots-disinformation.html}
  {Disinformation researchers raise alarms about a.i. chatbots}.
\newblock (Accessed: 19 Oct 2023).

\bibitem[{Hu et~al.(2023)Hu, Guo, Chen, Wen, and Yu}]{Hu:2023}
Xuming Hu, Zhijiang Guo, Junzhe Chen, Lijie Wen, and Philip~S. Yu. 2023.
\newblock \href {https://doi.org/10.1145/3539618.3591896} {Mr2: A benchmark for
  multimodal retrieval-augmented rumor detection in social media}.
\newblock In \emph{Proceedings of the 46th International ACM SIGIR Conference
  on Research and Development in Information Retrieval}, SIGIR '23, page
  2901–2912, New York, NY, USA. Association for Computing Machinery.

\bibitem[{Islam et~al.(2020{\natexlab{a}})Islam, Liu, Wang, and
  Xu}]{Rafiqul:2020b}
Md~Rafiqul Islam, Shaowu Liu, Xianzhi Wang, and Guandong Xu.
  2020{\natexlab{a}}.
\newblock Deep learning for misinformation detection on online social networks:
  a survey and new perspectives.
\newblock \emph{Social Network Analysis and Mining}, 10(1):1--20.

\bibitem[{Islam et~al.(2020{\natexlab{b}})}]{Islam:2020}
Md~Saiful Islam et~al. 2020{\natexlab{b}}.
\newblock \href {https://doi.org/10.4269/ajtmh.20-0812} {{COVID}-19–related
  infodemic and its impact on public health: A global social media analysis}.
\newblock \emph{The American Journal of Tropical Medicine and Hygiene},
  103(4):1621 -- 1629.

\bibitem[{Khattar et~al.(2019)Khattar, Goud, Gupta, and Varma}]{Khattar:2019}
Dhruv Khattar, Jaipal~Singh Goud, Manish Gupta, and Vasudeva Varma. 2019.
\newblock Mvae: Multimodal variational autoencoder for fake news detection.
\newblock In \emph{The World Wide Web Conference}, pages 2915--2921.

\bibitem[{Klie et~al.(2018)Klie, Bugert, Boullosa, Eckart~de Castilho, and
  Gurevych}]{Klie:2018}
Jan-Christoph Klie, Michael Bugert, Beto Boullosa, Richard Eckart~de Castilho,
  and Iryna Gurevych. 2018.
\newblock \href {https://aclanthology.org/C18-2002} {The {INCE}p{TION}
  platform: Machine-assisted and knowledge-oriented interactive annotation}.
\newblock In \emph{Proceedings of the 27th International Conference on
  Computational Linguistics: System Demonstrations}, pages 5--9. Association
  for Computational Linguistics.

\bibitem[{Krippendorff(1970)}]{Krippendorff:1970}
Klaus Krippendorff. 1970.
\newblock Estimating the reliability, systematic error and random error of
  interval data.
\newblock \emph{Educational and Psychological Measurement}, 30(1):61--70.

\bibitem[{Kwon and Cha(2014)}]{Kwon:2014}
Sejeong Kwon and Meeyoung Cha. 2014.
\newblock \href {https://ojs.aaai.org/index.php/ICWSM/article/view/14494}
  {Modeling bursty temporal pattern of rumors}.
\newblock \emph{Proceedings of the International AAAI Conference on Web and
  Social Media}, 8(1):650--651.

\bibitem[{Landis and Koch(1977)}]{Landis:1977}
J~Richard Landis and Gary~G Koch. 1977.
\newblock The measurement of observer agreement for categorical data.
\newblock \emph{Biometrics}, pages 159--174.

\bibitem[{Lee et~al.(2011)Lee, Eoff, and Caverlee}]{Lee:2011}
Kyumin Lee, Brian Eoff, and James Caverlee. 2011.
\newblock Seven months with the devils: A long-term study of content polluters
  on {T}witter.
\newblock In \emph{Proceedings of the International AAAI Conference on Web and
  Social Media}, pages 185--192.

\bibitem[{Li et~al.(2020)Li, Jiang, Shu, and Liu}]{Li:2020}
Yichuan Li, Bohan Jiang, Kai Shu, and Huan Liu. 2020.
\newblock \href {https://doi.org/10.1109/BigData50022.2020.9378472} {Toward a
  multilingual and multimodal data repository for {COVID}-19 disinformation}.
\newblock In \emph{2020 IEEE International Conference on Big Data (Big Data)},
  pages 4325--4330.

\bibitem[{Lucas et~al.(2022)Lucas, Cui, Le, and Lee}]{Lucas:2022}
Jason Lucas, Limeng Cui, Thai Le, and Dongwon Lee. 2022.
\newblock \href {https://doi.org/10.18653/v1/2022.constraint-1.11} {Detecting
  false claims in low-resource regions: A case study of {C}aribbean islands}.
\newblock In \emph{Proceedings of the Workshop on Combating Online Hostile
  Posts in Regional Languages during Emergency Situations}, pages 95--102,
  Dublin, Ireland. Association for Computational Linguistics.

\bibitem[{Ma et~al.(2017)Ma, Gao, and Wong}]{Ma:2017}
Jing Ma, Wei Gao, and Kam-Fai Wong. 2017.
\newblock \href {https://doi.org/10.18653/v1/P17-1066} {Detect rumors in
  microblog posts using propagation structure via kernel learning}.
\newblock In \emph{Proceedings of the 55th Annual Meeting of the Association
  for Computational Linguistics (Volume 1: Long Papers)}, pages 708--717,
  Vancouver, Canada. Association for Computational Linguistics.

\bibitem[{Mehta et~al.(2022)Mehta, Pacheco, and Goldwasser}]{Mehta:2022}
Nikhil Mehta, Maria~Leonor Pacheco, and Dan Goldwasser. 2022.
\newblock \href {https://doi.org/10.18653/v1/2022.acl-long.97} {Tackling fake
  news detection by continually improving social context representations using
  graph neural networks}.
\newblock In \emph{Proceedings of the 60th Annual Meeting of the Association
  for Computational Linguistics (Volume 1: Long Papers)}, pages 1363--1380,
  Dublin, Ireland. Association for Computational Linguistics.

\bibitem[{Memon and Carley(2020)}]{Memon:2020}
Shahan~Ali Memon and Kathleen~M Carley. 2020.
\newblock Characterizing {COVID}-19 misinformation communities using a novel
  {T}witter dataset.
\newblock \emph{arXiv preprint arXiv:2008.00791}.

\bibitem[{Morstatter et~al.(2014)Morstatter, Lubold, Pon-Barry, Pfeffer, and
  Liu}]{Morstatter:2014}
Fred Morstatter, Nichola Lubold, Heather Pon-Barry, J{\"u}rgen Pfeffer, and
  Huan Liu. 2014.
\newblock \href {https://doi.org/10.3115/v1/W14-2509} {Finding eyewitness
  tweets during crises}.
\newblock In \emph{Proceedings of the {ACL} 2014 Workshop on Language
  Technologies and Computational Social Science}, pages 23--27, Baltimore, MD,
  USA. Association for Computational Linguistics.

\bibitem[{Nakamura et~al.(2019)Nakamura, Levy, and Wang}]{Nakamura:2019}
Kai Nakamura, Sharon Levy, and William~Yang Wang. 2019.
\newblock r/fakeddit: A new multimodal benchmark dataset for fine-grained fake
  news detection.
\newblock \emph{arXiv preprint arXiv:1911.03854}.

\bibitem[{Neumann et~al.(2022)Neumann, De-Arteaga, and
  Fazelpour}]{Neumann:2022}
Terrence Neumann, Maria De-Arteaga, and Sina Fazelpour. 2022.
\newblock \href {https://doi.org/10.1145/3531146.3533205} {Justice in
  misinformation detection systems: An analysis of algorithms, stakeholders,
  and potential harms}.
\newblock In \emph{2022 ACM Conference on Fairness, Accountability, and
  Transparency}, FAccT '22, page 1504–1515, New York, NY, USA. Association
  for Computing Machinery.

\bibitem[{Nguyen et~al.(2020)Nguyen, Sugiyama, Nakov, and Kan}]{Nguyen:2020}
Van-Hoang Nguyen, Kazunari Sugiyama, Preslav Nakov, and Min-Yen Kan. 2020.
\newblock \href {https://doi.org/10.1145/3340531.3412046} {Fang: Leveraging
  social context for fake news detection using graph representation}.
\newblock In \emph{Proceedings of the 29th ACM International Conference on
  Information \& Knowledge Management}, CIKM '20, page 1165–1174, New York,
  NY, USA. Association for Computing Machinery.

\bibitem[{Nielsen and McConville(2022)}]{Nielsen:2022}
Dan~S. Nielsen and Ryan McConville. 2022.
\newblock \href {https://doi.org/10.1145/3477495.3531744} {Mumin: A large-scale
  multilingual multimodal fact-checked misinformation social network dataset}.
\newblock In \emph{Proceedings of the 45th International ACM SIGIR Conference
  on Research and Development in Information Retrieval}, SIGIR '22, page
  3141–3153, New York, NY, USA. Association for Computing Machinery.

\bibitem[{Nyhan and Reifler(2010)}]{Nyhan:2010}
Brendan Nyhan and Jason Reifler. 2010.
\newblock When corrections fail: The persistence of political misperceptions.
\newblock \emph{Political Behavior}, 32(2):303--330.

\bibitem[{Oshikawa et~al.(2020)Oshikawa, Qian, and Wang}]{Oshikawa:2020}
Ray Oshikawa, Jing Qian, and William~Yang Wang. 2020.
\newblock A survey on natural language processing for fake news detection.
\newblock In \emph{Proceedings of the 12th Language Resources and Evaluation
  Conference}, pages 6086--6093.

\bibitem[{Pan et~al.(2018)Pan, Pavlova, Li, Li, Li, and Liu}]{Pan:2018}
Jeff~Z Pan, Siyana Pavlova, Chenxi Li, Ningxi Li, Yangmei Li, and Jinshuo Liu.
  2018.
\newblock Content based fake news detection using knowledge graphs.
\newblock In \emph{International Semantic Web Conference}, pages 669--683.
  Springer.

\bibitem[{Paszke et~al.(2019)}]{Paszke:2019}
Adam Paszke et~al. 2019.
\newblock Pytorch: An imperative style, high-performance deep learning library.
\newblock In \emph{Advances in Neural Information Processing Systems 32}, pages
  8024--8035. Curran Associates, Inc.

\bibitem[{Pedregosa et~al.(2011)}]{Scikit:Learn}
F.~Pedregosa et~al. 2011.
\newblock Scikit-learn: Machine learning in {P}ython.
\newblock \emph{Journal of Machine Learning Research}, 12:2825--2830.

\bibitem[{P{\'e}rez-Rosas et~al.(2017)P{\'e}rez-Rosas, Kleinberg, Lefevre, and
  Mihalcea}]{Perez:2017}
Ver{\'o}nica P{\'e}rez-Rosas, Bennett Kleinberg, Alexandra Lefevre, and Rada
  Mihalcea. 2017.
\newblock Automatic detection of fake news.
\newblock \emph{arXiv preprint arXiv:1708.07104}.

\bibitem[{Reuters(2022)}]{Reuters:2022}
Reuters. 2022.
\newblock \href
  {https://www.reuters.com/world/europe/ukraine-russia-face-off-world-court-over-genocide-claim-2022-03-06/}
  {Russian no show at {U.N.} court hearing on ukrainian 'genocide'}.
\newblock (Accessed Oct 19, 2023).

\bibitem[{Rosenfeld et~al.(2020)Rosenfeld, Szanto, and Parkes}]{Rosenfeld:2020}
Nir Rosenfeld, Aron Szanto, and David~C. Parkes. 2020.
\newblock \href {https://doi.org/10.1145/3366423.3380180} {A kernel of truth:
  Determining rumor veracity on {T}witter by diffusion pattern alone}.
\newblock In \emph{Proceedings of The Web Conference 2020}, WWW '20, page
  1018–1028, New York, NY, USA. Association for Computing Machinery.

\bibitem[{Sch{\"u}tze et~al.(2008)Sch{\"u}tze, Manning, and
  Raghavan}]{Schutze:2008}
Hinrich Sch{\"u}tze, Christopher~D Manning, and Prabhakar Raghavan. 2008.
\newblock \emph{Introduction to information retrieval}, volume~39.
\newblock Cambridge University Press Cambridge.

\bibitem[{Schweter(2020)}]{Schweter:2020}
Stefan Schweter. 2020.
\newblock \href {https://doi.org/10.5281/zenodo.3770924} {{BERT}urk - {BERT}
  models for {T}urkish}.

\bibitem[{Shi and Weninger(2016)}]{Shi:2016}
Baoxu Shi and Tim Weninger. 2016.
\newblock \href {https://doi.org/https://doi.org/10.1016/j.knosys.2016.04.015}
  {Discriminative predicate path mining for fact checking in knowledge graphs}.
\newblock \emph{Knowledge-Based Systems}, 104:123--133.

\bibitem[{Shin et~al.(2018)Shin, Jian, Driscoll, and Bar}]{Shin:2018}
Jieun Shin, Lian Jian, Kevin Driscoll, and Fran{\c{c}}ois Bar. 2018.
\newblock The diffusion of misinformation on social media: Temporal pattern,
  message, and source.
\newblock \emph{Computers in Human Behavior}, 83:278--287.

\bibitem[{Shu et~al.(2019{\natexlab{a}})Shu, Bernard, and Liu}]{Shu:2019a}
Kai Shu, H~Russell Bernard, and Huan Liu. 2019{\natexlab{a}}.
\newblock Studying fake news via network analysis: detection and mitigation.
\newblock In \emph{Emerging Research Challenges and Opportunities in
  Computational Social Network Analysis and Mining}, pages 43--65. Springer.

\bibitem[{Shu et~al.(2020)Shu, Mahudeswaran, Wang, Lee, and Liu}]{Shu:2020}
Kai Shu, Deepak Mahudeswaran, Suhang Wang, Dongwon Lee, and Huan Liu. 2020.
\newblock \href {https://doi.org/10.1089/big.2020.0062} {Fakenewsnet: A data
  repository with news content, social context, and spatiotemporal information
  for studying fake news on social media}.
\newblock \emph{Big Data}, 8(3):171--188.

\bibitem[{Shu et~al.(2019{\natexlab{b}})Shu, Wang, and Liu}]{Shu:2019b}
Kai Shu, Suhang Wang, and Huan Liu. 2019{\natexlab{b}}.
\newblock \href {https://doi.org/10.1145/3289600.3290994} {Beyond news
  contents: The role of social context for fake news detection}.
\newblock In \emph{Proceedings of the Twelfth ACM International Conference on
  Web Search and Data Mining}, WSDM '19, page 312–320, New York, NY, USA.
  Association for Computing Machinery.

\bibitem[{Spitale et~al.(2023)Spitale, Biller-Andorno, and
  Germani}]{Giovanni:2023}
Giovanni Spitale, Nikola Biller-Andorno, and Federico Germani. 2023.
\newblock \href {https://doi.org/10.1126/sciadv.adh1850} {Ai model gpt-3
  (dis)informs us better than humans}.
\newblock \emph{Science Advances}, 9(26):eadh1850.

\bibitem[{Su et~al.(2020)Su, Wan, Liu, Huang et~al.}]{Su:2020}
Qi~Su, Mingyu Wan, Xiaoqian Liu, Chu-Ren Huang, et~al. 2020.
\newblock Motivations, methods and metrics of misinformation detection: An
  {NLP} perspective.
\newblock \emph{Natural Language Processing Research}, 1(1-2):1--13.

\bibitem[{Thirumuruganathan et~al.(2021)Thirumuruganathan, Simpson, and
  Lakshmanan}]{Thirumuruganathan:2021}
Saravanan Thirumuruganathan, Michael Simpson, and Laks~V.S. Lakshmanan. 2021.
\newblock \href {https://doi.org/10.1145/3448016.3452778} {To intervene or not
  to intervene: Cost based intervention for combating fake news}.
\newblock In \emph{Proceedings of the 2021 International Conference on
  Management of Data}, SIGMOD '21, page 2300–2309. Association for Computing
  Machinery.

\bibitem[{Toraman et~al.(2022{\natexlab{a}})Toraman, \c{S}ahinu\c{c}, and
  Yilmaz}]{Toraman:2022b}
Cagri Toraman, Furkan \c{S}ahinu\c{c}, and Eyup~Halit Yilmaz.
  2022{\natexlab{a}}.
\newblock \href {https://doi.org/10.1145/3501247.3531539} {Blacklivesmatter
  2020: An analysis of deleted and suspended users in {T}witter}.
\newblock In \emph{14th ACM Web Science Conference 2022}, WebSci '22, page
  290–295, New York, NY, USA. Association for Computing Machinery.

\bibitem[{Toraman et~al.(2022{\natexlab{b}})Toraman, Ozcelik,
  \c{S}ahinu{\c{c}}, and Sahin}]{Toraman:2022}
Cagri Toraman, Oguzhan Ozcelik, Furkan \c{S}ahinu{\c{c}}, and Umitcan Sahin.
  2022{\natexlab{b}}.
\newblock \href {http://ceur-ws.org/Vol-3180/paper-59.pdf} {{ARC-NLP} at
  checkthat!-2022: Contradiction for harmful tweet detection}.
\newblock In \emph{Proceedings of the Working Notes of {CLEF} 2022 - Conference
  and Labs of the Evaluation Forum, Bologna, Italy, September 5th - to - 8th,
  2022}, volume 3180 of \emph{{CEUR} Workshop Proceedings}, pages 722--739.

\bibitem[{Twitter(2022)}]{Twitter:2022b}
Twitter. 2022.
\newblock \href {https://twitter.github.io/birdwatch/} {Birdwatch is a
  collaborative way to add helpful context to tweets and keep people better
  informed}.
\newblock (Accessed Oct 19, 2023).

\bibitem[{Vapnik(1999)}]{Vapnik:1999}
Vladimir Vapnik. 1999.
\newblock \emph{The nature of statistical learning theory}.
\newblock Springer Science \& Business Media.

\bibitem[{Wang(2017)}]{Wang:2017}
William~Yang Wang. 2017.
\newblock \href {https://doi.org/10.18653/v1/P17-2067} {{``}{L}iar, liar pants
  on fire{''}: A new benchmark dataset for fake news detection}.
\newblock In \emph{Proceedings of the 55th Annual Meeting of the Association
  for Computational Linguistics (Volume 2: Short Papers)}, pages 422--426,
  Vancouver, Canada. Association for Computational Linguistics.

\bibitem[{Weinzierl and Harabagiu(2022)}]{Weinzierl:2022}
Maxwell Weinzierl and Sanda Harabagiu. 2022.
\newblock Vaccine{L}ies: A natural language resource for learning to recognize
  misinformation about the {COVID}-19 and {HPV} vaccines.
\newblock In \emph{Proceedings of the Language Resources and Evaluation
  Conference}, pages 6967--6975, Marseille, France. European Language Resources
  Association.

\bibitem[{Wolf et~al.(2020)}]{Wolf:2020}
Thomas Wolf et~al. 2020.
\newblock Transformers: State-of-the-art natural language processing.
\newblock In \emph{Proceedings of the 2020 Conference on Empirical Methods in
  Natural Language Processing: System Demonstrations}, pages 38--45, Online.
  Association for Computational Linguistics.

\bibitem[{Wu et~al.(2016)Wu, Morstatter, Hu, and Liu}]{Wu:2016}
Liang Wu, Fred Morstatter, Xia Hu, and Huan Liu. 2016.
\newblock Mining misinformation in social media.
\newblock In \emph{Big Data in Complex and Social Networks}, pages 135--162.
  CRC Press.

\bibitem[{Zellers et~al.(2019)Zellers, Holtzman, Rashkin, Bisk, Farhadi,
  Roesner, and Choi}]{Zellers:2020}
Rowan Zellers, Ari Holtzman, Hannah Rashkin, Yonatan Bisk, Ali Farhadi,
  Franziska Roesner, and Yejin Choi. 2019.
\newblock Defending against neural fake news.
\newblock In \emph{Advances in Neural Information Processing Systems},
  volume~32. Curran Associates, Inc.

\bibitem[{Zhou and Zafarani(2020)}]{Zhou:2020}
Xinyi Zhou and Reza Zafarani. 2020.
\newblock A survey of fake news: Fundamental theories, detection methods, and
  opportunities.
\newblock \emph{ACM Computing Surveys (CSUR)}, 53(5):1--40.

\bibitem[{Zhu et~al.(2023)Zhu, Zhang, Haq, Hui, and Tyson}]{Zhu:2023}
Yiming Zhu, Peixian Zhang, Ehsan-Ul Haq, Pan Hui, and Gareth Tyson. 2023.
\newblock \href {http://arxiv.org/abs/2304.10145} {Can chatgpt reproduce
  human-generated labels? a study of social computing tasks}.

\bibitem[{Zubiaga et~al.(2018)Zubiaga, Aker, Bontcheva, Liakata, and
  Procter}]{Zubiaga:2018}
Arkaitz Zubiaga, Ahmet Aker, Kalina Bontcheva, Maria Liakata, and Rob Procter.
  2018.
\newblock \href {https://doi.org/10.1145/3161603} {Detection and resolution of
  rumours in social media: A survey}.
\newblock \emph{ACM Comput. Surv.}, 51(2).

\end{thebibliography}

\end{document}